\documentstyle[12pt]{article}
\def\be{\begin{equation}}
\def\ee{\end{equation}}
\def\bea{\begin{eqnarray}}
\def\eea{\end{eqnarray}}

\begin{document}

\begin{center}
{\Large{\bf Tachyon Condensation on a Nonstationary D$p$-brane 
with Background Fields in Superstring Theory }}

\vskip .5cm
{\large F. Safarzadeh-Maleki and D. Kamani}
\vskip .1cm
{\it Physics department, Amirkabir University of Technology
(Tehran Polytechnic)}\\
P.O.Box: {\it 15875-4413, Tehran, Iran}\\
e-mails: {\sl f.safarzadeh@aut.ac.ir, kamani@aut.ac.ir}\\
\end{center}

\begin{abstract}

Using the boundary state formalism we obtain
the partition function corresponding to a dynamical
(rotating-moving) D$p$-brane in the presence of electromagnetic
and tachyonic background fields in the superstring theory.
The instability of such
D$p$-brane due to the tachyon condensation is investigated.

\end{abstract}



\vskip .5cm

\newpage

\section{Introduction}

D-branes can be described in
terms of closed string states, hence by using the boundary state
formalism many interesting properties have been
shown \cite{1}-\cite{9}. By means of the boundary state, all
relevant properties of the D-branes could be revealed. The
boundary state formalism has been applied to the various D-branes
configurations in the presence of different background fields
\cite{10}-\cite{14}.

On the other hand, investigating the stability of D-branes is one
of the most important subjects that can be studied via the
tachyon dynamics of open string and tachyon condensation
phenomenon \cite{15}. These concepts have been verified by various
methods \cite{16}-\cite{18} and more recently by the boundary
string field theory (BSFT) in different configurations
\cite{19}-\cite{24}. It has been conjectured that the open string
tachyon condensation describes the decay of unstable D-branes
into the closed string vacuum or to the lower dimensional unstable
D-branes as intermediate states. Study of this physical process
namely, decaying of unstable objects, is an important phenomenon
because of its interpolation between two different vacua and also
since it is a way to reach the concept of background independent
formulation of string theory.

Some aspects of the boundary state, accompanied by the tachyon
condensation, are as follow. The boundary state is a source for
closed strings, therefore, by using this state and tachyon
condensation, one can find the time evolution of the source for
each closed string mode. Also it has been argued that the
boundary state description of the rolling tachyon is valid during
the finite time which is determined by string coupling, and
the energy could be dissipated into the bulk beyond this time \cite{23}.
Moreover, this method shows the decoupling of the open string
modes at the non-perturbative minima of the tachyon potential
\cite{25}.

Previously we have calculated the boundary states associated with
a dynamical (rotating and moving) D$p$-brane in the presence of
the electromagnetic and tachyonic background fields \cite{12, 24}.
Now, by making use of the same boundary state we shall construct the
corresponding partition function, which is obtained by the BSFT
method. Then, we shall examine the instability of a
D$p$-brane. We demonstrate that this process can make such a
dynamical brane unstable, and hence reduces the brane's dimension.
\section{Boundary state of a dynamical brane}

For constructing a boundary state corresponding to a
dynamical (rotating-moving) D-brane
in the presence of some background fields, we start with the action
\bea
S &=& -\frac{1}{4\pi\alpha'} {\int}_\Sigma
d^{2}\sigma(\sqrt{-g}g^{ab}G_{\mu\nu}\partial_a X^{\mu}\partial_b
X^{\nu}+\varepsilon^{ab} B_{\mu\nu}\partial_a X^{\mu}\partial_b
X^{\nu})
\nonumber\\
&+& \frac{1}{2\pi\alpha'} {\int}_{\partial\Sigma} d\sigma ( A_\alpha
\partial_{\sigma}X^{\alpha}+ \omega_{\alpha\beta}J^{\alpha\beta}_{\tau}
+T(X^{\alpha})),
\eea
where $\Sigma$ and $\partial\Sigma$ are worldsheet of closed string
and its boundary, respectively.
This action contains the Kalb-Ramond field $B_{\mu\nu}$,
a $U(1)$ gauge field $A_\alpha$, an $\omega$-term
for rotation and motion of the brane and a tachyonic field.
We shall apply $\{X^\alpha|\alpha =0, 1, \cdot \cdot \cdot ,p \}$ for
the worldvolume directions of the brane and
$\{X^i| i= p+1, \cdot \cdot \cdot ,d-1\}$ for directions
perpendicular to it. 

The background fields $G_{\mu
\nu}$ and $B_{\mu \nu}$ are considered to be constant, and
for the $U(1)$ gauge field we use the gauge
$A_{\alpha}=-\frac{1}{2}F_{\alpha \beta }X^{\beta}$ which
possesses a constant field strength. Besides, the
tachyon profile
$T=\frac{1}{2}U_{\alpha\beta}X^{\alpha}X^{\beta}$ will
be used, where
the symmetric matrix $U_{\alpha\beta}$ is constant.
The $\omega$-term, which is responsible for the brane's
rotation and motion, contains the anti-symmetric angular velocity
${\omega }_{\alpha \beta}$ and angular momentum density
$J^{\alpha \beta }_{\tau}$ which is given by ${\omega }_{\alpha
\beta}J^{\alpha \beta }_{\tau}=2{\omega }_{\alpha \beta
}X^{\alpha }{\partial }_{\tau }X^{\beta }$. In fact, the component
$\omega_{0 {\bar \alpha}}|_{{\bar \alpha} \neq 0}$ denotes the
velocity of the brane along the direction $X^{\bar \alpha}$ while
$\omega_{{\bar \alpha}{\bar \beta}}$ represents its rotation.

It should be noted that rotation and motion of the brane are
considered to be in its volume.
In fact, according to the various fields inside the brane, the
Lorentz symmetry is broken and hence such a dynamic (rotation and
motion) is sensible.

Suppose that the following mixed elements vanish, i.e.
$B_{\alpha i} =U_{\alpha i} =0 $.
The oscillating part of the bosonic boundary state is given by
\bea
{|B_{\rm
Bos}\rangle}^{\left({\rm osc}\right)}\ =\prod^{\infty }_{n=1}
{[\det Q_{(n)}]^{-1}}\;{\exp \left[-\sum^{\infty }_{m=1}
{\frac{1}{m}{\alpha }^{\mu }_{-m}S_{(m)\mu \nu }
{\widetilde{\alpha }}^{\nu }_{-m}}\right]\ } {|0\rangle}_{\alpha}
\otimes {|0\rangle}_{\widetilde{\alpha }} \;,
\eea
in which the matrices are as follows:
\bea
&~& Q_{(n){\alpha \beta }} =
{\eta }_{\alpha \beta } -{{\mathcal F}}_{{\mathbf \alpha }{\mathbf
\beta }}+\frac{i}{2n}U_{\alpha \beta },
\nonumber\\
&~& S_{(m)\mu\nu}=(\Delta_{(m)\alpha \beta}\; ,\; -{\delta}_{ij}),
\nonumber\\
&~& \Delta_{(m)\alpha \beta} = (M_{(m)}^{-1}N_{(m)})_{\alpha
\beta},
\nonumber\\
&~& M_{(m){\alpha \beta }} = {\eta }_{\alpha \beta }+4{\omega
}_{\alpha \beta }-{{\mathcal F}}_{{\mathbf \alpha }{\mathbf \beta
}}+\frac{i}{2m}U_{\alpha \beta },
\nonumber\\
&~& N_{(m){\alpha \beta }} = {\eta }_{\alpha \beta } +4{\omega
}_{\alpha \beta } +{{\mathcal F}}_{{\mathbf \alpha }{\mathbf
\beta }} -\frac{i}{2m}U_{\alpha \beta },
\nonumber\\
&~& {\cal{F}}_{\alpha \beta}=\partial_\alpha A_\beta
-\partial_\beta A_\alpha - B_{\alpha \beta} .
\eea
The normalization
factor $\prod^{\infty }_{n=1}{{[\det Q_{(n){\alpha \beta
}}]}^{-1}}$ is an effect of the disk partition function.
In addition, the zero-mode part of the bosonic boundary state
has the feature
\bea
{{\rm |}B_{\rm
Bos}\rangle}^{\left(0\right)} &=& \frac{T_p}{2}\int^{\infty
}_{{\rm -}\infty } \exp\left\{i{\alpha }^{{\rm
'}}\left[\sum^{p}_{\alpha  =0} {\left(U^{{\rm -}{\rm 1}}{\mathbf
A}\right)}_{\alpha \alpha} {\left(p^{\alpha}\right)}^{{\rm
2}}{\rm +} \sum^{p}_{\alpha ,\beta {\rm =0},\alpha \ne
\beta}{{\left(U^{{\rm -}{\rm 1}}{\mathbf A}+{\mathbf A}^T
U^{-1}\right)}_{\alpha \beta } p^{\alpha
}p^{\beta}}\right]\right\}{\rm \ \ }
\nonumber\\
&\times& \left( \prod_{\alpha}{\rm |}p^{\alpha}\rangle
dp^{\alpha}\right) \otimes\prod_i{\delta {\rm (}x^i}{\rm
-}y^i{\rm )} {\rm |}p^i{\rm =0}\rangle ,
\eea
where
${\mathbf
A}_{\alpha \beta}=\eta_{\alpha \beta} + 4\omega_{\alpha \beta}$.

The NS-NS and R-R sectors possess the following fermionic 
boundary states
\bea 
&~& |B_{\rm Ferm} \rangle_{\rm NS}=\prod^{\infty}_{r=1/2}[\det
Q_{(r)}]\exp \bigg{[}i\sum^{\infty}_{r=1/2}(b^{\mu
}_{-r} S_{(r)\mu \nu}{\widetilde b}^{\nu}_{-r})\bigg{]}|0
\rangle ,\\
&~& |B_{\rm Ferm} \;\rangle_{\rm R}
=\prod^{\infty }_{n=1}[\det Q_{(n)}] {\exp
\left[i \sum^{\infty }_{m=1}{(d^{\mu }_{-m}S_{(m)\mu \nu }
{\widetilde{d}}^{\nu }_{-m})} \right]\ } |B\rangle^{(0)}_{\rm R}.
\eea
The explicit form of the zero-mode
state $|B\rangle^{(0)}_{\rm R}$ in the Type IIA and Type IIB
theories and its contribution to the spin structure can be found
in \cite{24} in complete details. It is not modified here because
for obtaining the partition function it will be projected onto
the bra-vacuum, hence, the remaining state would be the boundary
state built on the vacuum.

The total boundary state in the NS-NS and R-R sectors are
given by
\bea
|B \rangle_{\rm NS,R}=|B_{\rm Bos}\rangle^{({\rm osc})}
\otimes|B_{\rm Bos}\rangle^{(0)}
\otimes|B_{\rm Ferm}\rangle_{\rm NS,R} .
\eea
In fact, the total boundary state also has the ghosts and
superghosts boundary states. Since these parts are free
of the background fields, and specially free of 
the characteristic matrix of the tachyon, we put them away.
Note that the boundary state (7) contains
significant information about the nature of the brane.
\section{Tachyon condensation and collapse of a D$p$-brane}

The structure of the configuration space for the
boundary string field theory (BSFT) can be
described as follows: the space of 2-dimensional worldsheet theories on
the disk with arbitrary boundary interactions deals with the disk
partition function of the open string theory and a fixed
conformal worldsheet action in the bulk. It has been demonstrated
that, at the tree level, the disk partition function in the BSFT
appears as the normalization factor of the boundary state. In
other word, the partition function can be acquired by the vacuum
amplitude of the boundary state
\bea
Z^{\rm Disk}=\langle {\rm vacuum}|B\rangle.
\eea
Thus, in our setup the
partition function possesses the following feature
\bea
Z_{\rm Bos}^{\rm Disk}
&=& \frac{T_p}{2}\int^{\infty }_{{\rm -}\infty } {\prod_{\alpha
}{dp^{\alpha }}}\exp\bigg{\{}{i{\alpha }^{{\rm '}}\left[\sum^{p}_{\alpha =0}
{\left(U^{{\rm -}{\rm 1}}{\mathbf A}\right)}_{\alpha \alpha}
{\left(p^{\alpha}\right)}^{{\rm 2}}{\rm +} \sum^{p}_{\alpha
,\beta {\rm =0},\alpha \ne \beta}{{\left(U^{{\rm -}{\rm
1}}{\mathbf A}+{\mathbf A}^T U^{-1}\right)}_{\alpha \beta }
p^{\alpha }p^{\beta}}\right]\bigg{\}}{\rm \ \ }}
\nonumber\\
& \times & \prod^{\infty }_{n=1}{[\det Q_{(n)}]^{-1}}.
\eea
for the bosonic part of the partition function, and
\bea
Z_{\rm Ferm}^{\rm Disk}=\prod^{\infty}_{k>0}[\det Q_{(k)}],
\eea
for the fermionic part, where $k$ is half-integer (integer) for 
the NS-NS (R-R) sector. Therefore, after integrating on the momenta
and considering both fermionic and bosonic
parts, the total partition function in superstring theory is
given by 
\bea
Z_{\rm total}^{\rm Disk} &=& \frac{T_p}{2}
\left(\frac{i\pi}{\alpha'} \right)^{(p+1)/2}
\frac{1}{\sqrt{{\det (D +
H)}}}\frac{\prod^{\infty}_{k>0}[\det Q_{(k)}]}{\prod^{\infty
}_{n=1} {[\det Q_{(n)}]}},
\eea
where the diagonal matrix possesses the elements 
$D_{\alpha \beta}= (U^{-1}A)_{\alpha \alpha}\delta_{\alpha \beta}$, 
and the the matrix $H_{\alpha \beta}$ is defined by
\bea 
H_{\alpha \beta}= \bigg{\{} \begin{array}{c}
(U^{-1}A + A^T U^{-1})_{\alpha \beta}\;\;,\;\;\;\alpha \neq \beta ,\\
0 \;\;\;\;\;\;\;\;\;\;\;\;\;\;\;\;\;\;\;\;\;\;\;\;\;\;\;\;\;\;
\;,\;\;\;\alpha = \beta .
\end{array}
\eea 

The partition function enables us to investigate the 
effect of the tachyon
condensation on the instability of the D$p$-brane.
According to the conventional literature, the tachyonic mode of
open string spectrum makes the D-branes instable. This phenomenon
is called tachyon condensation. As the tachyon condenses, the
dimension of the brane decreases and in the final stage, one
receives a closed string vacuum. Using the boundary sigma-model,
the tachyon condensation usually starts with a conformal theory
with $d$ Neumann boundary conditions in the UV, and then adding
relevant tachyon field will cause the theory to roll toward an
IR fixed point as a closed string vacuum with a D$p$-brane, which
corresponds to a new vacuum with $(d-p-1)$ Dirichlet boundary
conditions.

According to the characteristic matrix of our tachyon, investigating
of the tachyon condensation in this work is more general than the
conventional studies which usually consider a single parameter
for the tachyon field. Now let's check the stability or instability
of the D$p$-brane in our setup. The tachyon condensation can be
occurred by taking at least one of the tachyon's elements to
infinity, i.e. $U_{pp} \to \infty$. At first look at the R-R
sector. By making use of the ${\mathop{\lim }_{U_{pp}\to \infty }
(U^{-1})_{p\alpha }=\mathop{{\rm \lim}}_{U_{pp}\to \infty }
(U^{-1})_{\alpha p}=0\ }$ the dimensional reduction of the
matrices $U^{-1}{\mathbf A}$, ${\mathbf A}^T U^{-1}$ and
$D$ is obvious. Therefore, according to Eq.
(11), in the R-R sector we observe that the direction $x^p$ has
been omitted from the resulted brane.

Now concentrate on the factor
$\prod^{\infty}_{r=\frac{1}{2}}[\det
Q_{(r)}]/\prod^{\infty }_{n=1} {[\det Q_{(n)}]}$ in the NS-NS
sector of the superstring partition function. Using the limit 
\bea
{{\mathop{\lim }_{U_{pp}\to \infty }\prod^{\infty}_{n=1} {{\bigg
[}\det {\bigg (}{\eta -{\cal{F}} +{\frac{iU}{2n}{\bigg )}}
_{(p+1)\times (p+1)}{\bigg ]}}^{-1}\ }\ }}
=\prod^{\infty}_{n=1}{\ \ \frac{2n}{iU_{pp}}\left[ \det{\left(
\eta -{\cal{F}} +\frac{iU}{2n}\right) }_{p\times p} \ \
\right]^{-1}} 
\eea 
the effect of tachyon condensation on this factor is given by 
\bea 
\mathop{\lim }_{U_{pp}\to \infty
}{\frac{\prod^{\infty}_{r=\frac{1}{2}}[\det (Q_{(r)})_{(p+1)\times
(p+1)}]}{\prod^{\infty }_{n=1} {[\det (Q_{(n)})_{(p+1)\times
(p+1)}]}}} \longrightarrow \sqrt{\frac{i\pi U_{pp}}{2}}\;
\frac{\prod^{\infty}_{r=\frac{1}{2}}[\det
(Q_{(r)})_{p\times p}]}{\prod^{\infty }_{n=1} {[\det
(Q_{(n)})_{p\times p}]}} 
\nonumber\\
\longrightarrow \sqrt{ \frac{i\pi U_{pp}}{2}\det(\eta -{\cal{F}})}\det
\left[\frac{\sqrt{\pi}\Gamma \left( 1+\frac{i}{2}(\eta
-{\cal{F}})^{-1}U \right)}{\Gamma \left(\frac{1}{2}+\frac{i}{2}(\eta
-{\cal{F}})^{-1}U \right) }\right]_{p\times p} . 
\eea 
The $p \times p$ matrices are similar to the initial 
$(p+1) \times (p+1)$ matrices in which the last rows and last 
columns have been omitted. In order to avoid
divergent quantities due to the existence of infinite product, 
in the second and third factors we
used the $\zeta$-function regularization. For this 
reason we used the arrow sign instead of equality.
However, it is evident that in this 
sector the dimensional reduction also occurs.

Let us check this factor after successive tachyon
condensation, i.e., 
\bea 
\mathop{\lim }_{U\to \infty
}{\frac{\prod^{\infty}_{r=\frac{1}{2}}[\det
(Q_{(r)})_{(p+1)\times (p+1)}]}{\prod^{\infty }_{n=1} {[\det
(Q_{(n)})_{(p+1)\times (p+1)}]}}}=\mathop{\lim }_{U\to \infty
}{\frac{\prod^{\infty}_{r=\frac{1}{2}}[\det
(\frac{iU}{2r})]}{\prod^{\infty }_{n=1} {[\det
(\frac{iU}{2n})]}}}
\longrightarrow
\mathop{\lim }_{U\to \infty
}\left(\frac{i\pi}{2}\right)^{(p+1)/2}\sqrt{\det U}\;, 
\eea 
where in the last term again, $\zeta$-function regularizations 
for infinite products have been used. Therefore, the total 
partition function finds the feature 
\bea 
Z_{\rm total}^{\rm Disk} &=& \frac{T_p}{2}
\left(-\frac{\pi^2}{2\alpha'} \right)^{(p+1)/2}
\sqrt{ \frac{\det U}{\det (D + H)}}\;.
\eea
In this limit, condensation would take place for all directions of the
brane's worldvolume. As can be seen, the dimensional reduction
followed by the sequential condensation process could not
disappear the tachyon.

For completing the discussion, let's see the tachyon
condensation effect via the boundary state approach, directly.
According to Eqs. (5) and (6), apart from
the normalization factors, i.e., the partition functions, look at
the $\Delta_{(m)}$ matrix in which the tachyon has been
entered. After applying the limit $U_{pp} \to \infty$,
this matrix possesses an eigenvalue ``-1'', i.e. 
we deduce that the Neumann direction $x^p$
has been omitted and instead it has been added to
the Dirichlet directions. This
process would be the same as in the bosonic case.

According to the above condensation processes, via the boundary
state and the BSFT approaches, the result is that 
in our setup the dimensional
reduction is taken place in both NS-NS and R-R sectors of the
superstring theory. That is, after tachyon
condensation, such a rotating-moving D$p$-brane with photonic and
tachyonic background fields, reduces to an unstable
D$(p-1)$-brane with its own background fields, rotation and
motion. Thus, imposing rotation and motion to an unstable
D-brane does not preserve it against collapse during the process
of tachyon condensation.

\end{document}